\documentclass[preprint,aps]{revtex4}

\usepackage{psfrag}
\usepackage{graphicx}
\usepackage{dcolumn}
\usepackage{color}
\usepackage{latexsym,amsfonts}
\usepackage{bm}
\usepackage{amssymb}
\begin{document}
\setcounter{page}{1}
\def\theequation{\arabic{section}.\arabic{equation}}
\def\theequation{\thesection.\arabic{equation}}
\setcounter{section}{0}

\title{Molecule model\\ for kaonic nuclear cluster $\bar{K}NN$}

\author{M. Faber${^a}$, A. N. Ivanov${^a}$, P. Kienle${^{b,c}}$,
J. Marton${^{b}}$, M. Pitschmann${^a}$}
\affiliation{${^a}$Atominstitut, Technische Universit\"at Wien,
Wiedner Hauptstrasse 8-10, A-1040 Wien, Austria}
\affiliation{${^b}$Stefan Meyer Institut f\"ur subatomare Physik
\"Osterreichische Akademie der Wissenschaften, Boltzmanngasse 3,
A-1090, Wien, Austria}\affiliation{${^c}$Excellence Cluster Universe
Technische Universit\"at M\"unchen, D-85748 Garching, Germany}
\email{ivanov@kph.tuwien.ac.at}

\date{\today}

\begin{abstract}
  We analyse the properties of the kaonic nuclear cluster (KNC)
  $\bar{K}NN$ with the structure $N\,\otimes (\bar{K}N)_{I = 0}$,
  having the quantum numbers $I(J^{\pi}) = \frac{1}{2}(0^-)$, and
  treated as a quasi--bound hadronic molecule state. We describe the
  properties of the hadronic molecule, or the KNC $N \otimes
  (\bar{K}N)_{I = 0}$, in terms of vibrational degrees of freedom with
  oscillator wave functions and chiral dynamics. These wave functions,
  having the meaning of trial wave functions of variational
  calculations, are parameterised by the frequency of oscillations of
  the $(\bar{K}N)_{I = 0}$ pair, which is fixed in terms of the
  binding energy of the strange baryon resonance $\Lambda(1405)$,
  treated as a quasi--bound $(\bar{K}N)_{I = 0}$ state. The binding
  energies $B_X$ and widths $\Gamma_X$ of the states $X =
  (\bar{K}N)_{I = 0}$ and $X = \bar{K}NN$, respectively, are
  calculated in the heavy--baryon approximation by using chiral
  Lagrangians with meson--baryon derivative couplings invariant under
  chiral $SU(3)\times SU(3)$ symmetry at the tree--level
  approximation. The results are $B_{\bar{K}NN} = 40.2\,{\rm MeV}$ and
  $\Gamma_{\bar{K}NN} = \Gamma^{(\not\pi)}_{\bar{K}NN} +
  \Gamma^{(\pi)}_{\bar{K}NN} \sim (85 - 106)\,{\rm MeV}$ and, where
  $\Gamma^{(\not\pi)}_{\bar{K}NN} \sim 21\,{\rm MeV}$ and
  $\Gamma^{(\pi)}_{\bar{K}NN} \sim (64 - 86)\,{\rm MeV}$ are the
  widths of non--pionic $\bar{K}NN \to N\Lambda^0, N \Sigma$ and
  pionic $\bar{K}NN \to N\Sigma\pi$ decay modes, calculated for
  $B_{\bar{K}N} = 29\,{\rm MeV}$ and $\Gamma_{\bar{K}N} = (30 -
  40)\,{\rm MeV}$, respectively. \\ PACS: 11.10.Ef, 13.75.Jz,
  24.10.Jv, 25.80.Nv
\end{abstract}  

\maketitle

\newpage

\section{Introduction}
\setcounter{equation}{0}

The kaonic nuclear cluster (KNC) $\bar{K}NN$ (or $K^-pp$) with quantum
numbers $I(J^{\pi}) = \frac{1}{2}(0^-)$ and the structure $N\otimes
(\bar{K}N)_{I = 0}$, which we denote as ${^2_{\bar{K}}}{\rm H}$
\cite{Akaishi1}, has been predicted by Akaishi and Yamazaki
\cite{Akaishi1}--\cite{Akaishi3} within a non--relativistic potential
model approach \cite{Akaishi1}--\cite{Yamazaki3} with the binding
energy $B_{{^2_{\bar{K}}}{\rm H}} = 48\,{\rm MeV}$ and the width
$\Gamma^{(\pi)}_{{^2_{\bar{K}}}{\rm H}} = 61\,{\rm MeV}$, taking into
account the pionic ${^2_{\bar{K}}}{\rm H} \to p \Sigma \pi$ decay
modes only \cite{Akaishi1}--\cite{Akaishi3}.  According to Akaishi and
Yamazaki \cite{Akaishi1}--\cite{Akaishi3}, the formation of the
quasi--bound $N\otimes (\bar{K}N)_{I = 0}$ state occurs in the
entrance channel through the isospin--singlet state $(\bar{K}N)_{I =
0}$, identified with the $\Lambda(1405)$ hyperon, with the subsequent
addition of one more proton.  In such a scenario the KNC
${^2_{\bar{K}}}{\rm H}$ acquires exactly the structure $N\, \otimes
(\bar{K}N)_{I = 0}$.

The theoretical analysis of the properties of the KNC
${^2_{\bar{K}}}{\rm H}$ is focused on the calculation of the binding
energy $B_{{^2_{\bar{K}}}{\rm H}}$ and the partial widths
$\Gamma^{(\not\pi)}_{{^2_{\bar{K}}}{\rm H}}$ and
$\Gamma^{(\pi)}_{{^2_{\bar{K}}}{\rm H}}$ of all hadronic decay modes,
which are the important two--body non--pionic ${^2_{\bar{K}}}{\rm H}
\to N\Lambda^0,N\Sigma$ decay modes and the three--body pionic
${^2_{\bar{K}}}{\rm H} \to N\Sigma\pi$ decay modes, respectively. In
the potential model approach by Akaishi and Yamazaki
\cite{Akaishi1}--\cite{Akaishi3} as well as in other theoretical
approaches for the description of the KNC ${^2_{\bar{K}}}{\rm H}$
\cite{KNC1}--\cite{KNC7} such as relativistic mean field (RMF) models
\cite{KNC1,KNC2}, coupled--channel approach with chiral dynamics and
variational analyses \cite{KNC3,KNC4}, coupled--channel Faddeev
equations \cite{KNC5,KNC6} and variational analyses \cite{KNC7} the
widths $ \Gamma^{(\pi)}_{{^2_{\bar{K}}}{\rm H}}$ of pionic modes
${^2_{\bar{K}}}{\rm H} \to N\Sigma\pi$ have been calculated only. Some
estimates of the non--pionic decay modes of the KNC
${^2_{\bar{K}}}{\rm H}$, carried out in the potential model approach
and coupled--channel approach with chiral dynamics, predict the widths
of order of $\Gamma^{(\not\pi)}_{{^2_{\bar{K}}}{\rm H}} \sim 10\,{\rm
MeV}$.  Recent calculation of the transitions $\Lambda^*N\to
N\Lambda^0$ and $\Lambda^*N\to N\Sigma$, where the $\Lambda(1405)$
resonance is denoted as $\Lambda^*$ \cite{Akaishi1}--\cite{Akaishi3},
which has been carried out in \cite{KNC8} at threshold of the unbound
$\Lambda^*N$ system, showed a total width of non--pionic decay modes
of the unbound $\Lambda^*N$ system equal to
$\Gamma^{(\not\pi)}_{\Lambda^*N} = 22\,{\rm MeV}$.

Since the decay mode ${^2_{\bar{K}}}{\rm H} \to p\Lambda^0$ is used
for the experimental observation of the KNC ${^2_{\bar{K}}}{\rm H}$
\cite{FINUDA,DISTO}, an alternative approach allowing a calculation of
the widths of all non--pionic and pionic decay modes for quasi--bound
${^2_{\bar{K}}}{\rm H}$ state is needed.

The first announcement about the existence of the KNC
${^2_{\bar{K}}}{\rm H}$ with the binding energy $B_{{^2_{\bar{K}}}{\rm
H}} = 115(7)\,{\rm MeV}$ and the total width
$\Gamma_{{^2_{\bar{K}}}{\rm H}} = 67(14)\,{\rm MeV}$ has been reported
by the FINUDA Collaboration \cite{FINUDA}. These data have been
obtained from the stopped $K^-$--meson reactions on ${^6}{\rm Li}$,
${^7}{\rm Li}$ and ${^{12}}{\rm C}$ by measuring the invariant-mass
spectrum of back--to--back emitted $p \Lambda^0$ pairs. However, the
interpretation of these results as a KNC ${^2_{\bar{K}}}{\rm H}$ has
been questioned and other explanations of the observed spectrum have
been put forward \cite{Ramos}.

Recent experimental data \cite{DISTO} on the analysis of the exclusive
$K^+$--meson missing mass and $p\Lambda^0$ invariant mass spectra of
the final state in the $pp \to K^+ \Lambda^0 p$ reaction at the
incident proton kinetic energy $T_p = 2.85\,{\rm GeV}$ showed the
quasi--bound state ${^2_{\bar{K}}}{\rm H}$ with the binding energy
$B_{{^2_{\bar{K}}}{\rm H}} = 103(6)\,{\rm MeV}$ and the width
$\Gamma_{{^2_{\bar{K}}}{\rm H}} = 118(13)\,{\rm MeV}$. The important
distinction of such a quasi--bound state is its two--body decays into
the $p\Lambda^0$ channel, enhanced at high momentum transfer and
predicted by the relevant reaction theory
\cite{Akaishi1}--\cite{Akaishi3}, showing a creation of a compact
object. Thus, for the analysis of such a quasi--bound state one cannot
use the coupled--channel Faddeev equation approach \cite{KNC5,KNC6},
which is unable to describe the two--body decay channels. In turn, a
description of the experimental data by the DISTO Collaboration within
other theoretical approaches \cite{KNC1}--\cite{KNC4,KNC7} seems to be
also questionable.

The KNC ${^2_{\bar{K}}}{\rm H}$ with the structure $N\,\otimes
(\bar{K}N)_{I = 0}$ we propose to treat as a kaonic molecule
\cite{Yamazaki1}.  In the center of mass frame molecular states are
defined by rotational and vibrational degrees of freedom
\cite{LL07}. Since the angular momentum of the KNC ${^2_{\bar{K}}}{\rm
H}$ as well as of the KNC ${^1_{\bar{K}}}{\rm H}$, the quasi--bound
$(\bar{K}N)_{I = 0}$ state, are equal to zero, the properties of these
states are defined by the vibrational degrees of freedom only, which
we describe by trial harmonic oscillator wave functions. Such a choice
can be also justified by a short--range character of forces producing
quasi--bound KNCs. We recall that the harmonic oscillator wave
functions are used also in the shell--model of nuclei \cite{Shell} for
the description of strongly bound nuclear systems. In addition, trial
Gaussian wave functions in the coordinate representation have been
used in \cite{KNC4} for the variational calculation of the parameters
of the KNC ${^2_{\bar{K}}}{\rm H}$ with a potential, induced by chiral
dynamics with $SU(3)$ coupled--channels technique.

The calculation of the parameters of the quasi--bound states
${^n_{\bar{K}}}{\rm H}$ with $n = 1,2,\ldots$ we propose to carry out
by using a $\mathbb{T}$--matrix, defined by chiral Lagrangians with
derivative meson--baryon couplings invariant under chiral $SU(3)\times
SU(3)$ symmetry \cite{ECL1,ECL2,CHPT}, which are used for the analysis
of low--energy strong interactions.

According to quantum field theoretic description of decays of
particles and nuclear states \cite{SS61}, the width of the decay modes
${^n_{\bar{K}}}{\rm H} \to X$ of the nuclear state ${^n_{\bar{K}}}{\rm
H}$ is defined by the decay amplitudes $M({^n_{\bar{K}}}{\rm H} \to
X)$ as
\begin{eqnarray}\label{label1.1}
  \hspace{-0.3in}\Gamma_{{^n_{\bar{K}}}{\rm H}} = \frac{1}{2
  M_{{^n_{\bar{K}}}{\rm H}}}\sum_X(2\pi)^4\delta^{(4)}(k_X -
  k_{{^n_{\bar{K}}}{\rm H}})|M({^n_{\bar{K}}}{\rm H} \to X)|^2 =
  \lim_{V,T\to \infty}\sum_X\frac{|\langle
  X|\mathbb{T}|{^2_{\bar{K}}}{\rm H}(\vec{0}\,) \rangle|^2 }{2
  M_{{^2_{\bar{K}}}{\rm H}}VT},
\end{eqnarray}
where we have used a relation $\langle X|\mathbb{T}|{^2_{\bar{K}}}{\rm
  H}(\vec{0}\,) = (2\pi)^4\delta^{(4)}(k_X - k_{{^n_{\bar{K}}}{\rm
  H}}) M({^n_{\bar{K}}}{\rm H} \to X)$. Then, $\sum_X$ assumes a
  summation over all allowed decay channels ${^n_{\bar{K}}}{\rm H} \to
  X$ and an integration over phase volumes of the final $X$--states,
  $|{^n_{\bar{K}}}{\rm H}(\vec{0}\,)\rangle$ is the wave function of
  the KNC ${^n_{\bar{K}}}{\rm H}$ in the momentum and particle number
  representation, the $\mathbb{T}$--matrix is defined by the chiral
  Lagrangians with derivative meson--baryon couplings invariant under
  non--linear chiral $SU(3)\times SU(3)$ transformations
  \cite{ECL1,ECL2}; $VT = (2\pi)^4 \delta^{(4)}(0)$ is a space--time
  volume \cite{SS61}.

Taking into account the unitarity condition for the
$\mathbb{T}$--matrix $\mathbb{T} - \mathbb{T}^{\dagger}=
i\mathbb{T}\mathbb{T}^{\dagger}$ we can generalise
Eq.(\ref{label1.1}) as follows
\begin{eqnarray}\label{label1.2}
  \hspace{-0.3in}&& B_{{^n_{\bar{K}}}{\rm H}} +\,i\,
  \frac{\Gamma_{{^n_{\bar{K}}}{\rm H}}}{2} = \lim_{V,T\to
  \infty}\frac{\langle{^n_{\bar{K}}}{\rm
  H}(\vec{0}\,)|\mathbb{T}|{^n_{\bar{K}}}{\rm H}(\vec{0}\,) \rangle
  }{2 M_{{^n_{\bar{K}}}{\rm H}}VT},
\end{eqnarray}
where $B_{{^n_{\bar{K}}}{\rm H}}$ is related to the energy shift of
the quantum state ${^n_{\bar{K}}}{\rm H}$ in the center of mass frame
as $B_{{^n_{\bar{K}}}{\rm H}} = -\,\epsilon_{{^n_{\bar{K}}}{\rm
H}}$. We identify it with the binding energy of the quantum state
${^n_{\bar{K}}}{\rm H}$. In such a definition the binding energy
$B_{{^n_{\bar{K}}}{\rm H}}$ and the width $\Gamma_{{^n_{\bar{K}}}{\rm
H}}$ of the KNC ${^n_{\bar{K}}}{\rm H}$ are determined by the same
low--energy strong interactions. This agrees with the definition of
the binding energy and the width of strongly coupled particles and
nuclear systems within the optical potential approach \cite{OPM}.

The width $\Gamma_{{^2_{\bar{K}}}{\rm H}}$ of the KNC
${^2_{\bar{K}}}{\rm H}$ is the sum of non--pionic decay modes
${^2_{\bar{K}}}{\rm H} \to NY$, where $NY = N\Lambda^0, N\Sigma^0$ and
$N\Sigma^+$, and pionic decay modes ${^2_{\bar{K}}}{\rm H} \to NY\pi$,
where $NY\pi = N \Sigma\pi$ and $N \Lambda^0 \pi$. It is given by
\begin{eqnarray}\label{label1.3}
  \hspace{-0.3in}\Gamma_{{^2_{\bar{K}}}{\rm H}} =
  \Gamma^{(\not\pi)}_{{^2_{\bar{K}}}{\rm H}} +
  \Gamma^{(\pi)}_{{^2_{\bar{K}}}{\rm H}} =
  \sum_{NY}\Gamma({^2_{\bar{K}}}{\rm H} \to NY) + \sum_{NY\pi}
  \Gamma({^2_{\bar{K}}}{\rm H} \to NY\pi).
\end{eqnarray}
In turn, the width $\Gamma_{{^1_{\bar{K}}}{\rm H}}$ of the KNC
${^1_{\bar{K}}}{\rm H}$ is defined by the ${^1_{\bar{K}}}{\rm H} \to
\Sigma\pi$ decay modes only as $\Gamma_{{^1_{\bar{K}}}{\rm H}} =
\Gamma({^1_{\bar{K}}}{\rm H} \to \Sigma\pi)$.

The paper is organised as follows.  In Section 2 we construct the wave
functions of the quasi--bound $(\bar{K}N)_{I = 0}$ and $N \otimes
(\bar{K}N)_{I = 0}$ states. Since angular momenta of these states are
zero, we describe them in terms of vibrational degrees of freedom by
the oscillator wave functions \cite{LL07,Shell} (see also
\cite{KNC4}).  We identify the quasi--bound $(\bar{K}N)_{I=0}$ state
with the strange baryon $\Lambda(1405)$, denoted below as $\Lambda^*$.
We give the analytical expressions for the binding energy and the
width of the quasi--bound $(\bar{K}N)_{I = 0}$ state in terms of the
frequency $\Omega_{\Lambda^*}$ of a motion of the $\bar{K}$ meson
relative to the nucleon $N$ and the analytical expression for the
binding energy of the quasi--bound $N \otimes (\bar{K}N)_{I = 0}$
state in terms of the frequencies $\Omega_{\Lambda^*}$ and
$\Omega_{\Lambda^*N}$, the later defining the motion of the
$(\bar{K}N)_{I = 0}$ system relative to the nucleon $N$. Thus, the
model contains three input parameters. They are the frequencies
$\Omega_{\Lambda^*}$ and $\Omega_{\Lambda^*N}$ and the coupling
constant $g_{\Lambda^*}$, which is used to fit the partial width of
the $\Lambda^*$ state to the values $\Gamma_{\Lambda^*} = (30 -
40)\,{\rm MeV}$ \cite{KNC8}. The number of the input parameters can be
reduced to $\Omega_{\Lambda^*}$ and $g_{\Lambda^*}$, which can be
obtained from the fit of the binding energy and width of the KNC
${^1_{\bar{K}}}{\rm H}$ only. For this aim we assume that the
stiffnesses of harmonic oscillator potentials, keeping the pairs
$(\bar{K}N)_{I = 0}$ and $N\otimes (\bar{K}N)_{I = 0}$ bound, are
equal. In this case the frequencies $\Omega_{\Lambda^*}$ and
$\Omega_{\Lambda^*N}$ become related by
$\mu_{\Lambda^*}\Omega^2_{\Lambda^*} =
\mu_{\Lambda^*N}\Omega^2_{\Lambda^*N}$, where $\mu_{\Lambda^*} =
m_Km_N/(m_N + m_K) = 324\,{\rm MeV}$ and $\mu_{\Lambda^*N} = m_N(m_N +
m_K)/(2 m_N + m_K) = 568\,{\rm MeV}$ are reduced masses of the
$\bar{K}N$ and $N(\bar{K}N)$ pairs, respectively. In Sections 3 we
give the analytical expressions for the amplitudes and widths of the
non--pionic $N \otimes (\bar{K}N)_{I = 0} \to NY$ and pionic $N
\otimes (\bar{K}N)_{I = 0} \to NY\pi$ decay modes of the quasi--bound
$N \otimes (\bar{K}N)_{I = 0}$ state. We make the calculations of the
binding energies and widths of the KNCs ${^1_{\bar{K}}}{\rm H}$ and
${^2_{\bar{K}}}{\rm H}$ at the tree--level approximation.  In this
case the binding energies $B_{{^1_{\bar{K}}}{\rm H}}$ and $
B_{{^2_{\bar{K}}}{\rm H}}$ of the KNCs ${^1_{\bar{K}}}{\rm H}$ and
${^2_{\bar{K}}}{\rm H}$ are induced by the Weinberg--Tomozawa
interactions only.  Such a dominance of the Weinberg--Tomozawa
low--energy strong interactions agrees well with the coupled--channel
approach, based on chiral dynamics \cite{KNC3,KNC4,L1420}. We denote
them as $B^{WT}_{{^1_{\bar{K}}}{\rm H}}$ and
$B^{WT}_{{^2_{\bar{K}}}{\rm H}}$. The masses of the KNCs
${^1_{\bar{K}}}{\rm H}$ and ${^2_{\bar{K}}}{\rm H}$ are given by
$M_{{^1_{\bar{K}}}{\rm H}} = m_N + m_K - B^{WT}_{{^1_{\bar{K}}}{\rm
H}}$ and $M_{{^2_{\bar{K}}}{\rm H}} = m_N + m_K -
B^{WT}_{{^2_{\bar{K}}}{\rm H}}$, respectively. In Section 4 we give
numerical values of the obtained binding energies and widths by using
for the analysis of the $\Lambda^*$ hyperon the following prediction
for its mass $m_{\Lambda^*} = 1405\,{\rm MeV}$ and widths
$\Gamma_{\Lambda^*} = 30\,{\rm MeV}$, obtained from the experimental
data on stopped--$K^-$ meson absorption in a deuteron target
\cite{L1405}, and $\Gamma_{\Lambda^*} = 40\,{\rm MeV}$, used by
Akaishi and Yamazaki in their original work \cite{Akaishi1}. In
Conclusion we discuss the obtained results.

\section{Quasi--bound $(\bar{K}N)_{I = 0}$ and $N \otimes (\bar{K}N)_{I
      = 0}$ states} \setcounter{equation}{0}

\subsection{Wave function of  quasi--bound $(\bar{K}N)_{I = 0}$ state}

The harmonic oscillator wave function of the KNC ${^1_{\bar{K}}}{\rm
H}$ in the momentum representation can be taken in the form
\cite{LL07}
\begin{eqnarray}\label{label2.1}
  \Phi_{{^1_{\bar{K}}}{\rm H}}(\vec{q}) =
  \Big(\frac{4\pi}{\mu_{\Lambda^*}\Omega_{\Lambda^*}}\Big)^{3/4}\,
\exp\Big(-\,\frac{\vec{q}^{\;2}}{2\mu_{\Lambda^*}\Omega_{\Lambda^*}}\Big),
\end{eqnarray} 
where $\Omega_{\Lambda^*}$ is the frequency of relative motion of the
$\bar{K}N$ pair.  In terms of the {\it stiffness} parameter $k$ the
frequency $\Omega_{\Lambda^*}$ is defined by $\Omega_{\Lambda^*} =
\sqrt{k/\mu_{\Lambda^*}}$, where $\mu_{\Lambda^*} = m_K m_N/(m_K +
m_N) = 324\,{\rm MeV}$ is the reduced mass of the $\bar{K}N$ pair,
calculated for $m_N = 940\,{\rm MeV}$ and $m_K = 494\,{\rm MeV}$.

In the particle number and momentum representation the wave function
of the KNC ${^1_{\bar{K}}}{\rm H}$ reads
\begin{eqnarray}\label{label2.2}
  \hspace{-0.3in}|{^1_{\bar{K}}}{\rm H}(\vec{k},\sigma)\rangle &=&
    \sqrt{2E_{{^1_{\bar{K}}}{\rm H}}
    (\vec{k}\,)}\int \frac{d^3k_1}{(2\pi)^3} \frac{d^3k_2}{(2\pi)^3}\,
    \frac{(2\pi)^3\delta^{(3)}(\vec{k} - \vec{k}_1 - \vec{k}_2)}{\sqrt{2
    E_{\bar{K}}(\vec{k}_1)2 E_N(\vec{k}_2)}}\nonumber\\
  \hspace{-0.3in}&&\times\, \Phi_{{^1_{\bar{K}}}{\rm H}}\Big(\frac{m_N
\vec{k}_1 - m_K\,\vec{k}_2}{m_K + m_N}\Big)\, \frac{1}{\sqrt{2}}
\sum_{j = 1,2}c^{\dagger}_{\bar{K}_j}(\vec{k}_1)\,
a^{\dagger}_{N_j}(\vec{k}_2,\sigma)|0\rangle,
\end{eqnarray}
where $\sigma = \pm\frac{1}{2}$ is a polarisation and $j$ is the
isospin index. The creation and annihilation operators obey standard
relativistic covariant commutation and anti--commutation relations
\cite{SS61}. The wave function (\ref{label2.2}) has a standard
relativistic covariant normalisation \cite{SS61}.

In our approach the binding energy and width of the KNC
${^1_{\bar{K}}}{\rm H}$ are defined by Eq.(\ref{label1.2}), where the
main contributions to the binding energies of quasi--bound
${^1_{\bar{K}}}{\rm H}$ and ${^2_{\bar{K}}}{\rm H}$ states are caused
by the Weinberg--Tomozawa low--energy strong interactions, which
produce the necessary attractions in the $(\bar{K}N)_{I = 0}$ and
$N\otimes (\bar{K}N)_{I = 0}$ systems allowing to treat them as
quasi--bound ${^1_{\bar{K}}}{\rm H}$ and ${^2_{\bar{K}}}{\rm H}$
states. This agrees well with the coupled--channel approach, based on
chiral dynamics \cite{KNC3,KNC4,L1420}. We perform the calculation of
the binding energy for the Weinberg--Tomozawa $(\bar{K}N)_{I = 0} \to
(\bar{K}N)_{I = 0}$ interaction and the width for the
Weinberg--Tomozawa--like $(\bar{K}N)_{I = 0} \to (\Sigma\pi)_{I = 0}$
interaction with a constant $g_{\Lambda^*}$, which we obtain from the
fit of the width of the KNC ${^1_{\!\bar{K}}}{\rm H}$. Such a
procedure of the calculation of the binding energy and width of the
KNC ${^1_{\bar{K}}}{\rm H}$ is equivalent to the fit of the optical
potential in the potential model approach by Akaishi and Yamazaki
\cite{Akaishi1}--\cite{Akaishi3}.  Then, we use these interactions for
the calculation of the binding energy and widths of the KNC
${^2_{\bar{K}}}{\rm H}$.

The calculation of the matrix elements of the $\mathbb{T}$--matrix in
the non--relativistic and heavy--baryon approximation gives the
following analytical expressions for the binding energy and the width
of the KNC ${^1_{\!\bar{K}}}{\rm H}$
\begin{eqnarray}\label{label2.3}
  B^{WT}_{{^1_{\!\bar{K}}}{\rm H}} &=&
  \frac{3}{4}\,\frac{1}{F^2_{\pi}}\Big\vert \int
  \frac{d^3q}{(2\pi)^3}\,\Phi_{{^1_{\!\bar{K}}}{\rm
  H}}(\vec{q}\,)\Big\vert^2 = \frac{3}{4}\,\frac{1}{F^2_{\pi}}\,
  \Big(\frac{\mu_{\Lambda^*}\Omega_{\Lambda^*}}{\pi}\Big)^{3/2},\nonumber\\
  \Gamma_{{^1_{\!\bar{K}}}{\rm H}} &=&
  g^2_{\Lambda^*}\frac{3}{8\pi}\,\frac{m_K
  m_{\Sigma}}{M_{{^1_{\!\bar{K}}}{\rm
  H}}}\,\frac{k_{\Sigma\pi}}{F^2_{\pi}}\,B_{{^1_{\!\bar{K}}}{\rm H}},
\end{eqnarray} 
where $g_{\Lambda^*}$ is a coupling constant of the
  ${^1_{\!\bar{K}}}{\rm H} \to \Sigma\pi$ decays,
  $M_{{^1_{\!\bar{K}}}{\rm H}} = m_K + m_p -
  B^{WT}_{{^1_{\!\bar{K}}}{\rm H}}$ is the mass of the quasi--bound
  ${^1_{\!\bar{K}}}{\rm H}$ state, $k_{\Sigma\pi} = 147\,{\rm MeV}$ is
  a relative momentum of the $\Sigma\pi$ pair and $m_{\Sigma} =
  1193\,{\rm MeV}$ and $m_{\pi} = 140\,{\rm MeV}$ are masses of the
  $\Sigma$--hyperon and the $\pi$--meson, respectively. The numerical
  values of $B^{WT}_{{^1_{\!\bar{K}}}{\rm H}}$ and
  $\Gamma_{{^1_{\!\bar{K}}}{\rm H}}$ are discussed in Section 4.

\subsection{Wave function of quasi--bound $N\otimes (\bar{K}N)_{I = 0}$ state}

The vibrational degrees of freedom of the molecule $N \otimes
(\bar{K}N)_{I = 0}$ are defined by the frequency $\Omega_{\Lambda^*}$,
the oscillations of the $\bar{K}N$ pair, and the frequency
$\Omega_{\Lambda^*N}$, the oscillations of the nucleon relative to the
$\bar{K}N$ pair.  The oscillator wave function of the KNC
${^2_{\bar{K}}}{\rm H}$, determined by $ \Phi_{{^2_{\bar{K}}}{\rm
H}}(\vec{k},\vec{k}_N) = \Phi_{\Lambda^*}(\vec{k}\,)
\Phi_{\Lambda^*N}(\vec{k}_N)$, is equal to
\begin{eqnarray}\label{label2.4}
 \hspace{-0.3in} \Phi_{{^2_{\bar{K}}}{\rm H}}(\vec{k},\vec{k}_N) =
\Big(\frac{4\pi}{\mu_{\Lambda^*}\Omega_{\Lambda^*}}\Big)^{3/4}
\Big(\frac{4\pi}{\mu_{\Lambda^*N} \Omega_{\Lambda^*N}}\Big)^{3/4}
\exp\Big(- \frac{\vec{k}^{\;2}}{2\mu_{\Lambda^*}\Omega_{\Lambda^*}} -
\frac{\vec{k}^{\;2}_N}{2\mu_{\Lambda^*N} \Omega_{\Lambda^*N}}\Big),
\end{eqnarray} 
where $\Omega_{\Lambda^*N} = \Omega_{\Lambda^*}\sqrt{\mu_{\Lambda^*}/
  \mu_{\Lambda^*N}}$ due to equal stiffnesses by assumption.

In the momentum and particle number representation the wave function
of the KNC ${^2_{\bar{K}}}{\rm H}$ with the structure $N \otimes
(\bar{K}N)_{I = 0}$ reads
\begin{eqnarray}\label{label2.5}
  \hspace{-0.3in}|{^2_{\bar{K}}}{\rm H}(\vec{k}\,)_j\rangle &=&
\sqrt{2E_{{^2_{\bar{K}}}{\rm H}}
(\vec{k}\,)}\int \frac{d^3k_1}{(2\pi)^3}
\frac{d^3k_2}{(2\pi)^3}\frac{d^3k_3}{(2\pi)^3} \,
\frac{\Phi_{{^2_{\bar{K}}}{\rm
H}}(\vec{k},\vec{k}_1,\vec{k}_2,\vec{k}_3)}{ \sqrt{2 E_N(k_1)
2E_N(k_2)2 E_{\bar{K}}(k_3) }}\nonumber\\
  \hspace{-0.3in}&&\times \frac{1}{\sqrt{2}}\sum_i
a^{\dagger}_{N_j}(\vec{k}_1,-\frac{1}{2})
a^{\dagger}_{N_i}(\vec{k}_2,+\frac{1}{2})c^{\dagger}_{\bar{K}_i}(\vec{k}_3)
|0\rangle,
\end{eqnarray}
where we have denoted
\begin{eqnarray}\label{label2.6}
   \Phi_{{^2_{\bar{K}}}{\rm H}}(\vec{k},\vec{k}_1,\vec{k}_2,\vec{k}_3)
   &=& (2\pi)^3\delta^{(3)}(\vec{k} - \vec{k}_1 - \vec{k}_2 -
   \vec{k}_3)\, \Phi_{\Lambda^*}\Big(\frac{m_K\vec{k}_2 - m_N
   \vec{k}_3}{m_K + m_N}\Big)\nonumber\\ &&\times\,
   \Phi_{\Lambda^*N}\Big(\frac{(m_K + m_N)\,\vec{k}_1 -
   m_N\,(\vec{k}_2 + \vec{k}_3)}{m_K + 2 m_N}\Big).
\end{eqnarray}
The wave function (\ref{label2.5}) has a standard relativistic
covariant normalisation \cite{SS61}. The wave functions
$\Phi_{\Lambda^*}$ and $\Phi_{\Lambda^*N}$ describe the two--body
$(\bar{K}N)_{I = 0}$ and three--body $N(\bar{K}N)_{I = 0}$
correlations of the $N\,\otimes (\bar{K}N)_{I = 0}$ molecule,
respectively. Of course, the wave function Eq.(\ref{label2.4}) takes
into account also the two--body $NN$ correlations.

\subsection{Binding energy of $N\otimes (\bar{K}N)_{I = 0}$ state }

The binding energy of the KNC ${^2_{\!\bar{K}}}{\rm H}$ is defined by
the attractive low--energy Weinberg--Tomozawa interactions with
isospin $I = 0$ and $I = 1$. The contribution of the
Weinberg--Tomozawa interaction with isospin $I=1$ appears due to $NN$
exchange interaction only.  The binding energy of the KNC
${^2_{\bar{K}}}{\rm H}$ is defined by
\begin{eqnarray}\label{label2.7}
 (B^{WT}_{{^2_{\!\bar{K}}}{\rm H}})_{j'j} = (B^{WT,I =
 0}_{{^2_{\!\bar{K}}}{\rm H}})_{j'j} + (B^{WT,I =
 1}_{{^2_{\!\bar{K}}}{\rm H}})_{j'j}.
\end{eqnarray}
The result of the calculation of the corresponding matrix elements of
the $\mathbb{T}$--matrix, obtained in the heavy--baryon approximation,
is
\begin{eqnarray}\label{label2.8}
  \hspace{-0.3in} (B^{WT}_{{^2_{\bar{K}}}{\rm H}})_{j'j} = \delta_{j'j}\,
\frac{3}{4}\,\frac{1}{F^2_{\pi}}\Bigg\{
\Big(\frac{\mu_{\Lambda^*}\Omega_{\Lambda^*}}{\pi}\Big)^{3/2} +
\frac{1}{2}\,\Big(\frac{\mu_{\Lambda^*N}
\Omega_{\Lambda^*N}}{\pi}\Big)^{3/2} \frac{1}{\displaystyle \Big(1 +
\frac{\mu_{\Lambda^*N}\Omega_{\Lambda^*N}}{\mu_{\Lambda^*}\Omega_{\Lambda^*}}\frac{\mu^2_{\Lambda^*}}{m^2_K}\Big)^{3/2}}\Bigg\}.
\end{eqnarray}
The contribution of the second term in (\ref{label2.8}) is caused by
the $NN$ exchange interaction and Weinberg--Tomozawa interactions with
isospin $I = 0$ and $I = 1$, respectively. It makes up of about
$24.7\,\%$ of the binding energy.
\begin{figure} \centering
\psfrag{K-}{$\bar{K}$} 
\psfrag{L0}{$\Lambda^0$} \psfrag{p0}{$\eta,\pi$}
\psfrag{p}{$N$}
 \psfrag{a}{$+$}
\psfrag{b}{$+ ~\ldots $}
\includegraphics[height= 0.12\textheight]{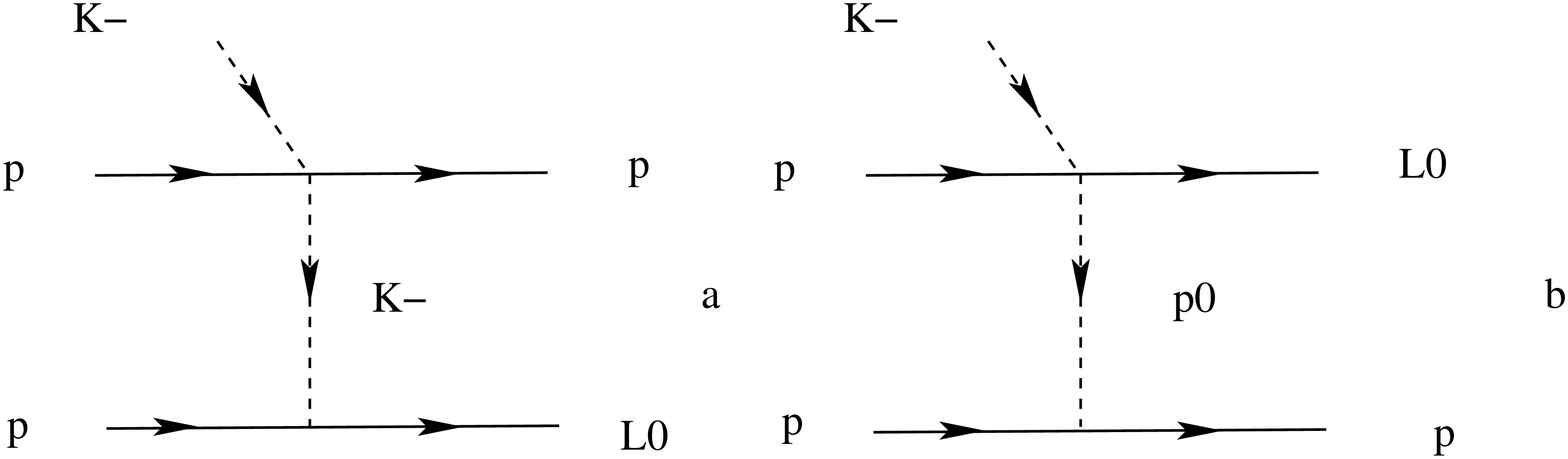}
\caption{The Feynman diagrams, defining the amplitude of the reaction
  $N(\bar{K}N)_{I = 0}\to N\Lambda^0$ in the molecule model of the KNC
  ${^2_{\bar{K}}}{\rm H}$. }
\end{figure}

\section{Decay modes of $N\otimes(\bar{K}N)_{I = 0}$ state}
\setcounter{equation}{0}

\subsection{Non--pionic decay modes of  $N\otimes(\bar{K}N)_{I = 0}$ state. 
Decay ${^2_{\bar{K}}}{\rm H} \to N\Lambda^0$ }

The amplitude of the $N(\bar{K}N)_{I = 0}\to N\Lambda^0$ reaction,
defining the non--pionic decay mode ${^2_{\bar{K}}}{\rm H} \to
N\Lambda^0$, is determined by the Feynman diagrams in Fig.\,1. The
analytical expression, obtained in the heavy--baryon approximation,
takes the form
\begin{eqnarray}\label{label3.1}
\hspace{-0.3in}&&M({^2_{\bar{K}}}{\rm H}_j \to
N_{j'}\Lambda^0) = \delta_{j'j}\,i\,\Big[\frac{9}{32}\frac{g_{\pi
NN}}{F^2_{\pi}}\Big]\,\sqrt{2 M_{{^2_{\bar{K}}}{\rm H}}
m_K}\,\Psi_{{^2_{\bar{K}}}{\rm H}}(0)\nonumber\\
\hspace{-0.3in}&&\times \Big\{\frac{4}{3}\,\frac{3 -
      2\alpha_D}{\sqrt{3}}\,\frac{1}{m^2_K + 2 T_{\Lambda^0}m_N} -
      \frac{3 - 4\alpha_D}{\sqrt{3}}\,\frac{1}{m^2_{\eta} + 2 T_N m_N}
      + \frac{1}{\sqrt{3}}\,\frac{1}{m^2_{\pi} + 2 T_N m_N}\Big\}\nonumber\\
\hspace{-0.3in}&&\times\, \frac{M_{{^2_{\bar{K}}}{\rm H}} +
      m_{\Lambda^0} + m_N}{\sqrt{2 m_{\Lambda^0}2
      m_N}}\,[\varphi^{\dagger}_{\Lambda^0}(\vec{\sigma}\cdot
      \vec{k}_{N\Lambda^0}\,)\chi_N] = \sqrt{2 M_{{^2_{\bar{K}}}{\rm
      H}}}\,{\cal M}({^2_{\bar{K}}}{\rm H}_j \to
N_{j'}\Lambda^0),
\end{eqnarray}
where $\varphi_{\Lambda^0}(\sigma_{\Lambda^0})$ and $\chi_N(\sigma_N)$
are spinorial wave functions of the $\Lambda^0$-- hyperon and the
nucleon, respectively, $g_{\pi NN} = g_A m_N/F_{\pi} = 13.3$ is the
$\pi NN$ coupling constant \cite{PNN}, $k_{N\Lambda^0}$ is a momentum
of a relative motion of the $N\Lambda^0$ pair, $T_{\Lambda^0}$ and
$T_N$ are kinetic energies of the baryons in the final $N\Lambda^0$
state.  Then, $M_{{^2_{\bar{K}}}{\rm H}} = 2 m_N + m_K -
B^{WT}_{{^2_{\bar{K}}}{\rm H}}$ is the mass of the KNC ${^2_{\bar{K}}}{\rm
H}$ and $\Psi_{{^2_{\bar{K}}}{\rm H}}(0)$ is the coordinate wave
function of the KNC ${^2_{\bar{K}}}{\rm H}$ at the origin
\begin{eqnarray}\label{label3.2}
 \Psi_{{^2_{\bar{K}}}{\rm H}}(0) =
\Psi_{\Lambda^*}(0)\Psi_{\Lambda^*N}(0) =
\Big(\frac{\mu_{\Lambda^*}\Omega_{\Lambda^*}}{\pi}
\frac{\mu_{\Lambda^*N}\Omega_{\Lambda^*N}}{\pi}\Big)^{3/4}.
\end{eqnarray}
The first and the second terms in the amplitude (\ref{label3.1}) are
defined by the $\bar{K}$--meson and $\eta$--meson exchange diagrams in
Fig.\,1, respectively, and Weinberg--Tomozawa interaction with isospin
$I = 0$. In turn, the contribution of the $\pi$--meson exchanges is
fully caused by the Weinberg--Tomozawa interaction with isospin $I =
1$ and the $NN$ exchange interaction.
\begin{figure} \centering
\psfrag{K-}{$\bar{K}$} 
\psfrag{L0}{$\Sigma$} \psfrag{p0}{$\pi,\eta$}
\psfrag{p}{$N$}
 \psfrag{a}{$+$}
\psfrag{b}{$+ ~\ldots $}
\includegraphics[height= 0.12\textheight]{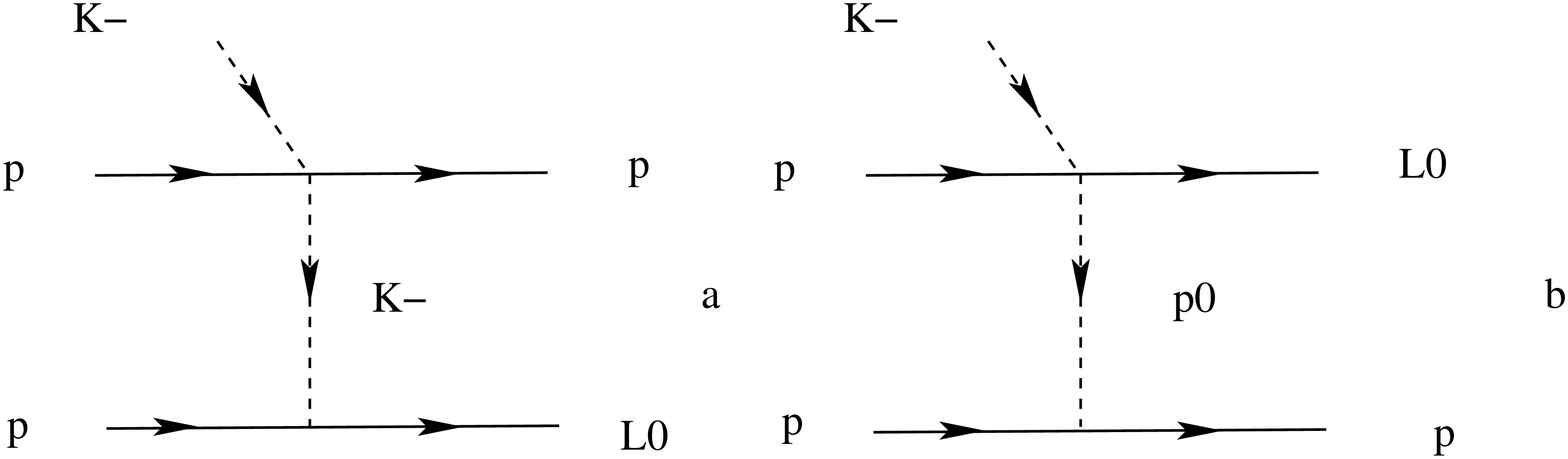}
\caption{The Feynman diagrams, defining the amplitudes of the
  reactions $N(\bar{K}N)_{I = 0}\to N\Sigma$ in the molecule model of
  the KNC ${^2_{\bar{K}}}{\rm H}$.}
\end{figure}
The structure
$[\varphi^{\dagger}_{\Lambda^0}(\vec{\sigma}\cdot
\vec{k}_{N\Lambda^0}\,)\chi_N]$ testifies that in the
${^2_{\bar{K}}}{\rm H} \to N \Lambda^0$ decay the $N\Lambda^0$ pair is
in the ${^3}{\rm P}_0$ state, whereas the KNC ${^2_{\bar{K}}}{\rm H}$
is in the ${^1}{\rm S}_0$ state. The partial width of the decay
${^2_{\bar{K}}}{\rm H}\to N\Lambda^0$ amounts to
\begin{eqnarray}\label{label3.3}
  \Gamma({^2_{\bar{K}}}{\rm H} \to N\Lambda^0) = \frac{1}{4\pi}
  \sum_{\sigma_{\Lambda^0}, \sigma_N = \pm\frac{1}{2}}
  |{\cal M}({^2_{\bar{K}}}{\rm H} \to N\Lambda^0)|^2\,
  \frac{|\vec{k}_{N\Lambda^0}|}{M_{{^2_{\bar{K}}}{\rm H}}},
\end{eqnarray}
where we have summed over polarisations of the particles in the final
state. The numerical value of $\Gamma({^2_{\bar{K}}}{\rm H} \to
N\Lambda^0)$ is discussed in Section 4.

\subsection{Non--pionic decay modes of  $N\otimes(\bar{K}N)_{I = 0}$ state. 
Decay ${^2_{\bar{K}}}{\rm H} \to N\Sigma$ }

The amplitudes of the $N(\bar{K}N)_{I = 0}\to N\Sigma$ reactions,
defining the non--pionic decay modes ${^2_{\bar{K}}}{\rm H} \to
N\Sigma$, are determined by the Feynman diagrams in Fig.\,2. The
analytical expression, obtained in the heavy--baryon approximation,
takes the form
\begin{eqnarray}\label{label3.4}
\hspace{-0.3in}&&M({^2_{\bar{K}}}{\rm H}_j \to
N_{j'}\Sigma^a) = -i\,(\tau^a)_{j'j}\,\Big[\frac{9}{32}\frac{g_{\pi
NN}}{F^2_{\pi}}\Big]\,\sqrt{2 M_{{^2_{\bar{K}}}{\rm H}}
m_K}\,\Psi_{{^2_{\bar{K}}}{\rm H}}(0)\nonumber\\
\hspace{-0.3in}&&\times \Big\{ (2\alpha_D - 1)\,\Big(g_{\Lambda^*} -
\frac{1}{9}\Big)\,\frac{1}{m^2_K + 2 T_{\Sigma}m_N} - \Big(
\frac{1}{\sqrt{3}} +\frac{2}{9}\Big)\,\frac{1}{m^2_{\pi} + 2 T_N
  m_N}\nonumber\\
\hspace{-0.3in}&& -\frac{3 - 4\alpha_D}{9}\,\frac{1}{m^2_{\eta} + 2
      T_N m_N}\Big\}\,\frac{M_{{^2_{\bar{K}}}{\rm H}} + m_{\Sigma} +
      m_N}{\sqrt{ 2 m_{\Sigma}2
      m_N}}\,[\varphi^{\dagger}_{\Sigma}(\vec{\sigma}\cdot
      \vec{k}_{N\Sigma})\chi_N] = \nonumber\\
\hspace{-0.3in}&&= \sqrt{2 M_{{^2_{\bar{K}}}{\rm H}}}\,{\cal
      M}({^2_{\bar{K}}}{\rm H}_j \to N_{j'}\Sigma^a),
\end{eqnarray}
where $g_{\Lambda^*}$ is the coupling constant of the
${^1_{\bar{K}}}{\rm H} \to \Sigma\pi$ decays. The contributions of the
Weinberg--Tomozawa interactions with isospin $I = 1$ survive due to
the $NN$ exchange interactions for the $\bar{K}$, $\pi$ and $\eta$
meson exchanges. The partial width of the ${^2_{\bar{K}}}{\rm H}\to
N\Sigma^0$ decay amounts to
\begin{eqnarray}\label{label3.5}
  \Gamma({^2_{\bar{K}}}{\rm H} \to N\Sigma^0) = \frac{1}{4\pi}
  \sum_{\sigma_{\Lambda^0}, \sigma_N = \pm\frac{1}{2}}
  |{\cal M}({^2_{\bar{K}}}{\rm H} \to N\Sigma)|^2\,
  \frac{|\vec{k}_{N\Sigma}|}{M_{{^2_{\bar{K}}}{\rm H}}}.
\end{eqnarray}
The partial widths of the decay modes ${^2_{\bar{K}}}{\rm H}\to
N\Sigma^0$ and ${^2_{\bar{K}}}{\rm H}\to N\Sigma^+$ are related by
$\Gamma({^2_{\bar{K}}}{\rm H} \to N\Sigma^+) = 2
\Gamma({^2_{\bar{K}}}{\rm H} \to N\Sigma^0)$. The numerical values are
discussed in Section 4.

\subsection{Pionic decay modes of  $N\,\otimes (\bar{K}N)_{I =
    0}$ state}

Since the $\bar{K}N$ pair in the KNC ${^2_{\bar{K}}}{\rm H}$ is in the
isospin--singlet state, the dominant pionic three--body decay modes
are ${^2_{\bar{K}}}{\rm H} \to N\Sigma\pi $.  In the
tree--approximation the Feynman diagram of the decay modes
${^2_{\bar{K}}}{\rm H} \to N\Sigma\pi $ is shown in Fig.\,3.
\begin{figure}
\centering
\psfrag{K-}{$\bar{K}$} 
\psfrag{L}{$\Lambda^*$} 
\psfrag{S}{$\Sigma$} 
\psfrag{p}{$N$}
\psfrag{pi}{$\pi$}
\psfrag{b}{$+ ~\ldots$}
\includegraphics[height= 0.15\textheight]{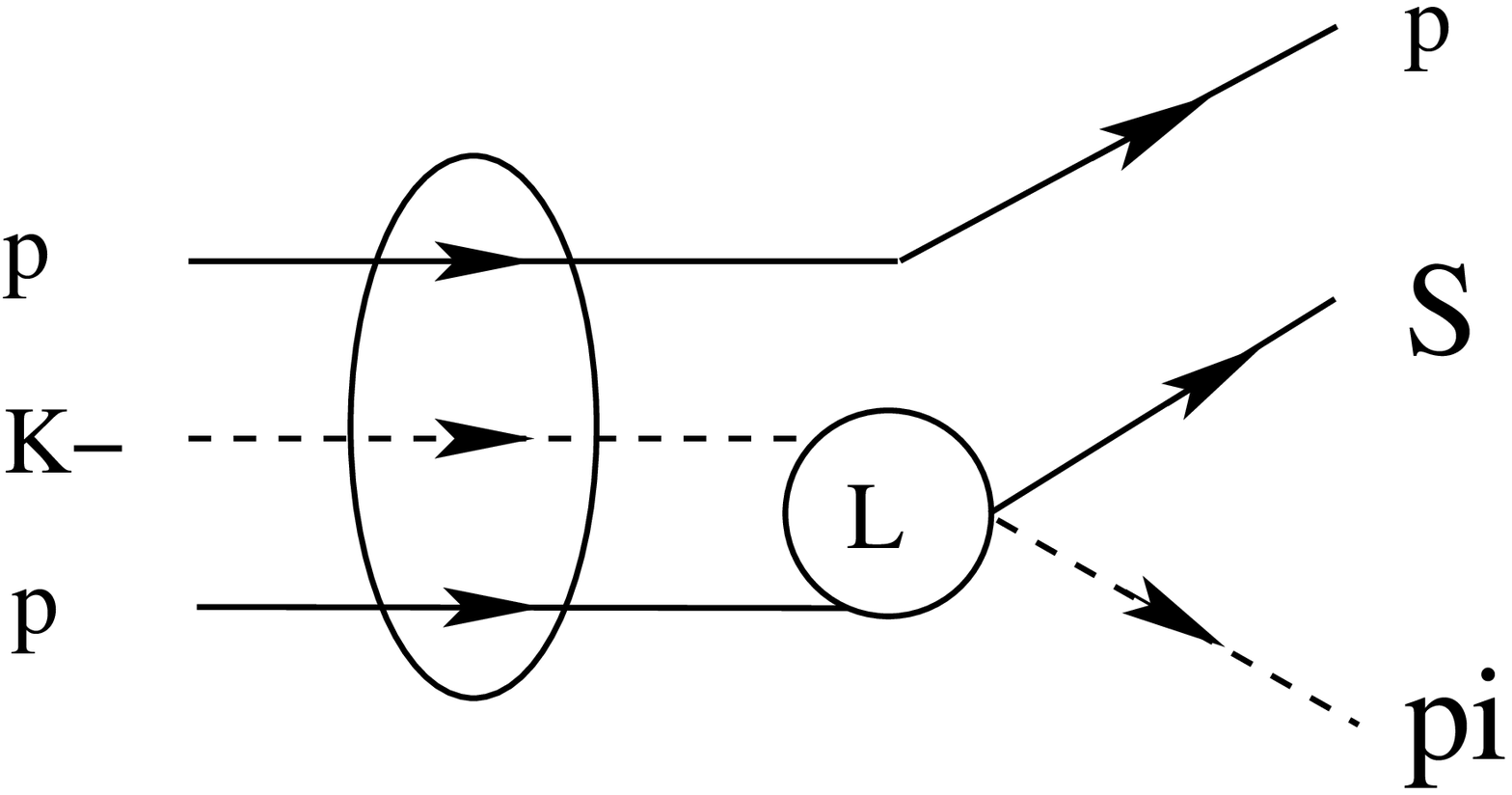}
\caption{The Feynman diagram of the decay mode ${^2_{\bar{K}}}{\rm
    H}\to N\Sigma \pi$ of the KNC ${^2_{\bar{K}}}{\rm H}$.}
\end{figure}
The amplitude of the ${^2_{\bar{K}}}{\rm H}\to N\Sigma \pi$ decay is
equal to
\begin{eqnarray}\label{label3.6}
\hspace{-0.3in}M({^2_{\bar{K}}}{\rm H}_j \to N_{j'}\Sigma^a\pi^b) &=&
-\,\delta_{j'j}\,\delta^{ab}\,\frac{1}{2}\,\frac{g_{\Lambda^*}}{F^2_{\pi}}\,
\sqrt{6 M_{{^2_{\bar{K}}}{\rm H}} m_{\Sigma}
  m_Nm_K}\,\Psi_{\Lambda^*}(0)\Phi_{\Lambda^*N}(0)\nonumber\\
\hspace{-0.3in}&&\times\,{\cal
  M}(\Omega_{\Lambda^*},k^2_N)_{\sigma_N,\sigma_{\Sigma}},
\end{eqnarray}
where $k_N$ is a 3---momentum of the nucleon, $\sigma_{\Sigma} = \pm
1/2$ and $\sigma_N = \pm 1/2$ are polarisations of the
$\Sigma$--hyperon and the nucleon in the final state, respectively,
$\Psi_{\Lambda^*}(0) = (\mu_{\Lambda^*}\Omega_{\Lambda^*}/\pi)^{3/4}$
and $\Phi_{\Lambda^*N}(0) =
(4\pi/\mu_{\Lambda^*N}\Omega_{\Lambda^*N})^{3/4}$ and ${\cal
M}(\Omega_{\Lambda^*},k^2_N)_{\sigma_N,\sigma_{\Sigma}}$ is equal to
\begin{eqnarray}\label{label3.7}
\hspace{-0.3in}&&{\cal
  M}(\Omega_{\Lambda^*},k^2_N)_{\sigma_N,\sigma_{\Sigma}} =
  \Bigg\{\Phi_{\Lambda^*N}(\vec{k}_N) + \frac{1}{B_{{^2_{\bar{K}}}{\rm
  H}} - B_{{^1_{\bar{K}}}{\rm
  H}} + \vec{k}^{\,2}_N/2\mu_{\Lambda^*N}}\nonumber\\
\hspace{-0.3in}&&\times\,\frac{3}{4}\,\frac{1}{F^2_{\pi}}\Bigg(
  \Phi_{\Lambda^*N}(\vec{k}_N)|\Psi_{\Lambda^*}(0)|^2 -
  \frac{1}{2}\Psi_{\Lambda^*}(0)\int\frac{d^3q}{(2\pi)^3}
  \Phi_{\Lambda^*}\Big(\vec{k}_N + \frac{\mu_{\Lambda^*}}{
  m_N}\,\vec{q}\,\Big)\Phi_{\Lambda^*N}(\vec{q}\,)\nonumber\\
\hspace{-0.3in}&& + \frac{3}{4} \int
    \frac{d^3k}{(2\pi)^3}\frac{d^3q}{(2\pi)^3}
    \,\Phi^*_{\Lambda^*}\Big(\vec{k} +
    \frac{\mu_{\Lambda^*}}{m_K}\,(\vec{q}+
    \vec{k}_N)\Big)\Phi_{\Lambda^*}(\vec{k}\,)\Phi_{\Lambda^*N}(\vec{q}\,)
    \Bigg)\Bigg\}\, \delta_{\sigma_N,-\frac{1}{2}}
    \delta_{\sigma_{\Sigma},+\frac{1}{2}}\nonumber\\
\hspace{-0.3in}&&- \Bigg\{\frac{1}{2}\int
\frac{d^3q}{(2\pi)^3}\,\Phi^*_{\Lambda^*}\Big(\vec{q} +
\frac{\mu_{\Lambda^*}}{
m_K}\,\vec{k}_N\Big)\,\Phi_{\Lambda^*}\Big(\vec{k}_N +
\frac{\mu_{\Lambda^*}}{
m_K}\,\vec{q}\,\Big)\,\Phi_{\Lambda^*N}(\vec{q}\,) + \frac{1}{
B^{WT}_{{^2_{\bar{K}}}{\rm H}} - B^{WT}_{{^1_{\bar{K}}}{\rm H}} +
\vec{k}^{\,2}_N/2\mu_{\Lambda^*N}}\nonumber\\
\hspace{-0.3in}&&\times\,
\frac{3}{4}\,\frac{1}{F^2_{\pi}}\Bigg(\frac{1}{4}
\int\frac{d^3k}{(2\pi)^3}\frac{d^3q}{(2\pi)^3}\,\Phi^*_{\Lambda^*}\Big(\vec{k}
+ \frac{\mu_{\Lambda^*}}{m_K}(\vec{q} +
\vec{k}_N)\Big)\Phi_{\Lambda^*}(\vec{k}\,)\Phi_{\Lambda^*N}(\vec{q}\,)\nonumber\\
\hspace{-0.3in}&&- \frac{1}{2}\Psi_{\Lambda^*}(0)
\int\frac{d^3q}{(2\pi)^3}\,\Phi^*_{\Lambda^*}\Big(\vec{q} +
\frac{\mu_{\Lambda^*}}{m_K }
\vec{k}_N\Big)\Phi_{\Lambda^*N}(\vec{q}\,)\Bigg)\Bigg\} \,
\delta_{\sigma_N,+\frac{1}{2}}\delta_{\sigma_{\Sigma},-\frac{1}{2}}.
\end{eqnarray}
The amplitude of the ${^2_{\bar{K}}}{\rm H}\to N\Sigma \pi$ decay is
defined by the Weinberg--Tomozawa--like interaction, responsible for
the ${^1_{\bar{K}}}{\rm H}\to \Sigma \pi$ decay, and the
Weinberg--Tomozawa interactions $(\bar{K}N)_{I=0} \to
(\bar{K}N)_{I=0}$ and $(\bar{K}N)_{I=1} \to (\bar{K}N)_{I=1}$ with
isospin $I = 0$ and $I = 1$, respectively.  The contribution of the
Weinberg--Tomozawa interaction $(\bar{K}N)_{I=1} \to (\bar{K}N)_{I=1}$
with isospin $I = 1$ appears due to the $NN$ exchange interaction. The
partial width of the ${^2_{\bar{K}}}{\rm H}\to N \Sigma \pi$ decays is
\begin{eqnarray}\label{label3.8}
\Gamma({^2_{\bar{K}}}{\rm H} \to N\Sigma \pi) =
\frac{9}{4}\frac{g^2_{\Lambda^*}}{F^4_{\pi}}
|\Psi_{\Lambda^*}(0)|^2|\Phi_{\Lambda^*N}(0)|^2\,
M^2_{{^2_{\bar{K}}}{\rm H}}m_{\Sigma}m_N m_K
f_{N\Sigma\pi}(\Omega_{\Lambda^*}),
\end{eqnarray}
where $f_{N\Sigma\pi}(\Omega_{\Lambda^*})$, caused by the contribution
of the phase volume of the final $N\Sigma\pi$ state, is defined by
\begin{eqnarray}\label{label3.9}
\hspace{-0.3in}&&f_{p\Sigma\pi}(\Omega_{\Lambda^*}) =
\frac{1}{128\pi^3 M^4_{{^2_{\bar{K}}}{\rm H}}}
\int^{(M_{{^2_{\bar{K}}}{\rm H}} - m_N)^2}_{(m_{\Sigma} +
m_{\pi})^2}\frac{ds}{s}\sqrt{(s - (m_{\Sigma} + m_{\pi})^2)(s -
(m_{\Sigma} - m_{\pi})^2} \nonumber\\
\hspace{-0.3in}&&\times \sqrt{(M_{{^2_{\bar{K}}}{\rm H}} + m_N)^2 -
s)(M_{{^2_{\bar{K}}}{\rm H}} - m_N)^2 - s}\sum_{\sigma_N =
\pm\frac{1}{2}}\sum_{\sigma_{\sigma_{\Sigma}}= \pm\frac{1}{2} }|{\cal
M}(\Omega_{\Lambda^*},k^2_N)_{\sigma_N,\sigma_{\Sigma}}|^2.
\end{eqnarray}
\begin{table}[h]
\begin{tabular}{|c|c|c|c|}
\hline & \multicolumn{2}{|c|}{Molecule model} & Potential Model \\
\hline & $M_{{^1_{\bar{K}}}{\rm H}} = 1405\,{\rm
MeV}$&$M_{{^1_{\bar{K}}}{\rm H}} = 1405\,{\rm MeV}$
&$M_{{^1_{\bar{K}}}{\rm H}} = 1405\,{\rm MeV}$ \\\hline
$B^{WT}_{{^1_{\bar{K}}}{\rm H}}$ & $29.0\,{\rm MeV}$ & $29.0\,{\rm
MeV}$ &$27.0\,{\rm MeV}$\\ \hline $\Gamma_{{^1_{\bar{K}}}{\rm H}}$ &
$30.0\,{\rm MeV}$& $40.0\,{\rm MeV}$ &$40.0\,{\rm MeV}$\\\hline
$\Omega_{\Lambda^*}$& $46.3\,{\rm MeV}$ & $46.3\,{\rm MeV}$ &\\ \hline
$g_{\Lambda^*}$& $1.095$ & $1.265$ &\\ \hline $\Omega_{\Lambda^*N}$ &
$35.0\,{\rm MeV}$ &$35.0\,{\rm MeV}$ & \\\hline
$B^{WT}_{{^2_{\bar{K}}}{\rm H}}$ & $40.2\,{\rm MeV}$ &$40.2\,{\rm
MeV}$ &$48\,{\rm MeV}$ \\ \hline $\Gamma({^2_{\bar{K}}}{\rm H} \to
N\Lambda^0)$ & $14.6\,{\rm MeV}$ &$14.6\,{\rm MeV}$ &\\ \hline
$\Gamma({^2_{\bar{K}}}{\rm H} \to N\Sigma^0)$ & $2.2\,{\rm MeV}$ &
$2.0\,{\rm MeV}$&\\ \hline $\Gamma({^2_{\bar{K}}}{\rm H} \to
N\Sigma^+)$ & $4.4\,{\rm MeV}$ &$4.0\,{\rm MeV}$ &\\\hline
$\Gamma^{(\not\pi)}_{{^2_{\bar{K}}}{\rm H}}$ & $21.2\,{\rm MeV}$ &
$20.6\,{\rm MeV}$ &$\approx 12\,{\rm MeV}$\\ \hline
$\Gamma^{(\pi)}_{{^2_{\bar{K}}}{\rm H}} $ & $64.0\,{\rm MeV}$
&$85.5\,{\rm MeV}$ &$61\,{\rm MeV}$\\ \hline
$\Gamma_{{^2_{\bar{K}}}{\rm H}} =
\Gamma^{(\not\pi)}_{{^2_{\bar{K}}}{\rm H}} +
\Gamma^{(\pi)}_{{^2_{\bar{K}}}{\rm H}} $ & $85.2\,{\rm MeV}$
&$106.1\,{\rm MeV}$ &$ \approx 73\,{\rm MeV}$\\\hline
\end{tabular} 
\caption{The binding energies and widths of the KNC
${^2_{\bar{K}}}{\rm H}$}
\end{table}
The numerical value of the partial width of the pionic decay modes we
discuss in Section 4.

\section{Numerical values  of binding energy and widths  of 
${^2_{\bar{K}}}{\rm H}$} 
\setcounter{equation}{0}

In this section we give numerical values of the binding energy and
partial widths of the KNC ${^2_{\bar{K}}}{\rm H}$ for the KNC
${^1_{\bar{K}}}{\rm H}$ with mass $M_{{^1_{\bar{K}}}{\rm H}} =
1405\,{\rm MeV}$ and widths $\Gamma_{{^1_{\bar{K}}}{\rm H}} = 30\,{\rm
MeV}$ and $\Gamma_{{^1_{\bar{K}}}{\rm H}} = 40\,{\rm MeV}$, predicted
by potential model \cite{L1405} and \cite{Akaishi1}, respectively. In
Table I we summarise the results of our molecule model and the
potential model.

One can see that the binding energy $B_{^2_{\bar{K}}{\rm H}} =
40.2\,{\rm MeV}$ and the width $\Gamma_{^2_{\bar{K}}{\rm H}} =
85.2\,{\rm MeV}$ of the KNC ${^2_{\bar{K}}}{\rm H}$, calculated for
the KNC ${^1_{\bar{K}}}{\rm H}$ with mass $M_{{^1_{\bar{K}}}{\rm H}} =
1405\,{\rm MeV}$ and the width $\Gamma_{{^1_{\bar{K}}}{\rm H}} =
30\,{\rm MeV}$, agree well with the results, obtained in the potential
model approach by Akaishi and Yamazaki.  The width of the non--pionic
decay modes $\Gamma^{(\not\pi)}_{{^1_{\bar{K}}}{\rm H}} \simeq
21\,{\rm MeV}$ agrees well with the width
$\Gamma^{(\not\pi)}_{\Lambda^*N} = 22\,{\rm MeV}$ of the unbound
$\Lambda^*N$ state, calculated recently in \cite{KNC8}. Our results
agree also well with those obtained within the coupled--channel
Faddeev equation approach \cite{KNC5,KNC6}. The agreement with other
approaches \cite{KNC1}-\cite{KNC4,KNC7} is only qualitative.

\section{Conclusion}
\setcounter{equation}{0}

We have investigated the properties of the simplest kaonic nuclear
clusters (KNCs) ${^1_{\bar{K}}}{\rm H}$ and ${^2_{\bar{K}}}{\rm H}$
with the structures $(\bar{K}N)_{I= 0}$ and $N\,\otimes (\bar{K}N)_{I
= 0}$, respectively, in the model, which we call ``Molecule model for
kaonic nuclear clusters''. It is based on the assumption that KNCs are
hadronic molecules \cite{Yamazaki1}. In our model the calculation of
the binding energies and the widths of KNCs is a kind of variational
calculation with trial wave functions taken in the form of harmonic
oscillator wave functions. Such a choice is justified as follows.  It
is known \cite{LL07} that molecules are described by rotational and
vibrational degrees of freedom. Since rotational degrees of freedom
are absent for the KNC ${^1_{\bar{K}}}{\rm H}$ and ${^2_{\bar{K}}}{\rm
H}$, their properties are determined by vibrational degrees of freedom
only. The binding energy and the width of the KNCs ${^1_{\bar{K}}}{\rm
H}$ and ${^2_{\bar{K}}}{\rm H}$ are defined by the diagonal elements
$\langle{^1_{\bar{K}}}{\rm H}|\mathbb{T}|{^1_{\bar{K}}}{\rm H}\rangle$
and $\langle{^2_{\bar{K}}}{\rm H}|\mathbb{T}|{^2_{\bar{K}}}{\rm
H}\rangle$ of the $\mathbb{T}$--matrix, which are calculated by using
chiral Lagrangians with derivative meson--baryon couplings invariant
under chiral $SU(3)\times SU(3)$ symmetry at the tree--level
approximation and in the heavy--baryon approximation.  The dominant
contributions to the binding energies and the widths come from the
Weinberg--Tomozawa interactions. This agrees well with $SU(3)$
coupled--channel approach with chiral dynamics.

The KNC $(\bar{K}N)_{I = 0}$ is identified with the hyperon
$\Lambda(1405)$, the wave function of which is taken in the form of
the trial harmonic oscillator wave function with a frequency
$\Omega_{\Lambda^*}$, describing the relative motion or correlations
of the $\bar{K}N$ pair in the state with $I = 0$.  The wave function
of the KNC $N\,\otimes (\bar{K}N)_{I = 0}$ is defined by the
frequencies $\Omega_{\Lambda^*}$ and $\Omega_{\Lambda^*N}$, where the
latter describes a motion of a nucleon $N$ relative to the $\bar{K}N$
pair in the state with $I = 0$.  They take into account two--body
$\bar{K}N$ and three--body $N(\bar{K}N)$ correlations in the KNC
${^2_{\bar{K}}}{\rm H}$, respectively, and, of course, the two--body
$NN$ correlations.

The calculations of the binding energy and widths of the KNC
${^2_{\bar{K}}}{\rm H}$ have been carried out for the low--lying KNC
${^1_{\bar{K}}}{\rm H}$ with mass $M_{{^1_{\bar{K}}}{\rm H}} =
1405\,{\rm MeV}$ and widths $\Gamma_{{^1_{\bar{K}}}{\rm H}} = (30 -
40)\,{\rm MeV}$, predicted by the potential model approach from the
experimental data on the stopped--$K^-$ meson absorption in the
deuteron target \cite{L1405} and used by Akaishi and Yamazaki in their
original work \cite{Akaishi1}, respectively.  Recently, the
theoretical analysis of the contribution of the $\Lambda(1405)$
resonance to the cross sections for elastic and inelastic $K^-p$
scattering and elastic $\pi\Sigma \to \pi \Sigma$ scattering has been
carried out in \cite{JR09}. Following the results obtained in
\cite{JR09}, one can conclude that in the pure $I = 0$ channel $\pi^0
\Sigma^0\to \pi^0 \Sigma^0$ the maximum of the cross section is
located around $m_{\Lambda^*} \simeq 1405\,{\rm MeV}$. It has been
also found experimentally \cite{COSY} that the shape and position of
the $\Lambda(1405)$ distribution, reconstructed in the $\Sigma^0\pi^0$
channel, are consistent with mass $m_{\Lambda^*} \sim 1400\,{\rm MeV}$
and width $\Gamma_{\Lambda^*} \sim 60\,{\rm MeV}$, agreeing well with
the shape and position of the $\Lambda(1405)$ distribution, measured
in the charge--exchange channels \cite{RH84,RD91}.

We have shown that treating the $\Lambda(1405)$ resonance as a
quasi--bound $(\bar{K}N)_{I = 0}$ with the mass, defined by
$M_{{^1_{\bar{K}}}{\rm H}} = m_K + m_N - B^{WT}_{{^1_{\bar{K}}}{\rm
H}}$, and the widths $\Gamma_{\Lambda^*} = (30 - 40)\,{\rm MeV}$, the
molecule model for kaonic nuclear clusters describes the KNC
${^2_{\bar{K}}}{\rm H}$ in a qualitative agreement with Akaishi and
Yamazaki \cite{Yamazaki1} and the results, obtained in the
coupled--channel Faddeev equation approach \cite{KNC5,KNC6}.

Our results for the widths of non--pionic decay modes
$\Gamma^{(\not\pi)}_{{^2_{\bar{K}}}{\rm H}} \simeq 21\,{\rm MeV}$
agree well with the result $\Gamma^{(\not\pi)}_{\Lambda^*N} = 22\,{\rm
MeV}$, obtained in \cite{KNC8}. Unlike \cite{KNC8} our analysis of the
non--pionic decay modes ${^2_{\bar{K}}}{\rm H} \to N\Lambda^0$ and
${^2_{\bar{K}}}{\rm H} \to N\Sigma$ takes into account the $NN$
exchange interactions, which play an important role for the correct
description of the properties of the KNC ${^2_{\bar{K}}}{\rm H}$.

The explanation of the experimental data by the DISTO Collaboration
$B_{{^2_{\bar{K}}}{\rm H}} = 103(6)\,{\rm MeV}$ and
$\Gamma_{{^2_{\bar{K}}}{\rm H}} = 118(13)\,{\rm MeV}$ \cite{DISTO} in
the molecule model for kaonic nuclear clusters is possible, but it
goes beyond the description of the KNCs ${^1_{\bar{K}}}{\rm H}$ and
${^2_{\bar{K}}}{\rm H}$ at the tree--level approximation for the
binding energies and the assumption of the equal stiffnesses of
harmonic oscillator potentials. The analysis of the experimental data
by the DISTO Collaboration in the molecule model for kaonic nuclear
clusters we carry out in the forthcoming paper.

\section{Acknowledgement}

The work has been delivered in part at the Workshop on ``Hadronic
Atoms and Kaonic Nuclei - solved puzzles, open problems and future
challenges in theory and experiment'' held on 12 - 16 October 2009 at
European Centre for Theoretical Studies in Nuclear Physics and Related
Areas (${\rm E^*CT}$) in Trento, Italy.

We are grateful to A. Gal for fruitful comments and discussions during
the Workshop and J. R$\acute{\rm e}$vai for discussions of the problem
of the $\Lambda(1405)$ resonance.

This research was partly supported by the DFG cluster of excellence
"Origin and Structure of the Universe" of the Technische Universit\"at
M\"unchen and by the Austrian ``Fonds zur F\"orderung der
Wissenschaftlichen Forschung'' (FWF) under contract P19487-N16.


\begin{thebibliography}{9}
\bibitem{Akaishi1} 
Y. Akaishi and T. Yamazaki, Phys. Rev. C {\bf 65},
044005 (2002); T. Yamazaki and Y. Akaishi, Phys. Lett. B {\bf 535}, 70
(2002).
\bibitem{Yamazaki1} 
T. Yamazaki and Y. Akaishi, Proc. Japan. Acad. B
{\bf 83}, 144 (2007); arXiv: 0706.3651 [nucl--th];
{\rm http://www.jstage.jst.go.jp/browse/pjab}
\bibitem{Yamazaki2}
T. Yamazaki and Y. Akaishi,
Nucl. Phys. A {\bf 792}, 229 (2007).
\bibitem{Yamazaki3}
T. Yamazaki and Y. Akaishi,
Phys. Rev. C {\bf 76}, 045201 (2007).
\bibitem{Akaishi2}
Y. Akaishi,
Mod. Phys. Lett. A {\bf 23}, 2516 (2008).
\bibitem{Akaishi3}
Y. Akaishi and T. Yamazaki,
Int. J. Mod. Phys. A {\bf 24}, 2118 (2009).
\bibitem{KNC1}
A. Gal, Nucl. Phys. A {\bf 790}, 143 (2007).
\bibitem{KNC2}
J. Mare${\check{\rm s}}$, E. Friedman, and A. Gal,
Int. J. Mod. Phys. A {\bf 22}, 633 (2007);
D. Gazda, E. Friedman, A. Gal, and J. Mare${\check{\rm s}}$,
Phys. Rev. C {\bf 76}, 055204 (2007). 
\bibitem{KNC3}
A. Dot$\acute{\rm e}$, T. Hyodo,  and W. Weise,
Nucl. Phys. A {\bf 804}, 197 (2008).
\bibitem{KNC4}
A. Dot${\acute{\rm e}}$, T. Hyodo,  and W. Weise, 
Phys. Rev. C {\bf 79}, 014003 (2009).
\bibitem{KNC5}
N. V. Shevchenko, A. Gal, and J. Mare${\check{\rm s}}$,
Phys. Rev. Lett. {\bf 98}, 082301 (2007).
N. V. Shevchenko, A. Gal, J. Mare${\check{\rm s}}$, and J. R$\acute{\rm e}$vai,
Phys. Rev. C {\bf 76}, 044004 (2007).
\bibitem{KNC6}
Y. Ikeda and T. Sato,
Phys. Rev. C {\bf 76}, 035203 (2007).
\bibitem{KNC7}
S. Wycech and A. M. Green,
Int. J. of Mod. Phys. A {\bf 22}, 629 (2007);
Phys. Rev. C {\bf 79}, 014001 (2009). 
\bibitem{KNC8}
T. Sekihara, D. Jido, and Y. Kanada--En'yo,
Phys. Rev. C {\bf 79}, 062201(R) (2009).
\bibitem{FINUDA}
M. Agnello {\it et al.} (FINUDA) Collaboration), Phys. Rev. Lett. {\bf 94},
212303 (2005).
\bibitem{DISTO} 
T. Yamazaki {\it et al.}, (DISTO Collaboration),
Phys. Rev. Letters {\bf 104}, 132502 (2010).
\bibitem{Ramos}
A. Ramos {\it et al.},
Eur. Phys. J. A {\rm 31}, 684 (2007).
\bibitem{LL07} 
L. D. Landau and E. M. Lifschitz, in {\it
Quantenmechanik, Nichtrelativistische N\"aherung}, Lehrbuch der
Theoretischen Physik, Band III, Verlag Harri Deutsch, 2007.
\bibitem{Shell}
A. deShalit and H. Feshbach,
in {\it Theoretical nuclear physics}, Vol. I {\it Nuclear structure},
John $\&$ Sons, Inc., New York, 1974.
\bibitem{ECL1}
B. W. Lee,
Phys. Rev. {\bf 170}, 1359 (1968).
\bibitem{ECL2}
B. Borasoy, R. Ni\ss ler, and W. Weise, 
Eur. Phys. J. A
  {\bf 25}, 79 (2005).
\bibitem{CHPT}
J. Gasser, V. E. Lyubovitskij, and  A. Rusetsky
Phys. Rep. {\bf 456}, 167 (2008) and references therein.
\bibitem{SS61} 
S. S. Schweber, in {\it An introduction to relativistic
    quantum field theory}, Row, Peterson and Co$\,\bullet\,$ Evanston,
  Ill., Elmsford, New York, 1961.
\bibitem{OPM}
P. B. Jones,
in {\it The optical model in nuclear and particle physics},
Interscience Publishers, John $\&$ Sons, New York, 1963;
H. Lenske and P. Kienle,
Phys. Lett. B {\bf 647}, 82 (2007) and references therein.
\bibitem{L1405}
J. Esmaili, Y. Akaishi, and T. Yamazaki,
arXiv: 0909.2573 [nucl--th].
\bibitem{L1420}
T. Hyodo and W. Weise,
Phys. Rev. C {\bf 77}, 035204 (2008).
\bibitem{PDG08}
C. Amsler {\it et al.},
Phys. Lett. B {\bf 667}, 1 (2008).
\bibitem{PNN}
T. E. O. Ericson, B. Loiseau, and S. Wycech,
Phys. Lett. B {\bf 594}, 76 (2004).
\bibitem{JR09}
J. R${\acute{\rm e}}$vai and N. Shevchenko,
Phys. Rev. C {\bf 79}, 035202 (2009).
\bibitem{COSY}
I Zychor {\it at el.},
Phys. Lett. B {\bf 660}, 167 (2008).
\bibitem{RH84}
 R. J. Hemingway, Nucl. Phys. B {\bf 253}, 742 (1984).
\bibitem{RD91}
R. H. Dalitz, A. Deloff, J. Phys. G {\bf 17}, 289  (1991).
\end{thebibliography}
\end{document}